\documentclass[12pt,a4j]{article}

\usepackage[dvipdfmx]{graphicx}

\makeatletter
\def\slash#1{{\mathpalette\c@ncel{#1}}} 
\makeatother

\newcommand\beq{\begin{eqnarray}}
\newcommand\eeq{\end{eqnarray}}

\newcommand\la{\langle}
\newcommand\ra{\rangle}

\def\Sslash{\rlap/{\mkern-1mu S}}
\def\pslash{\rlap/{\mkern-1mu p}}

\def\Sslash{\slash{\mkern-1mu S}}

\begin{document}

\begin{flushright}
\end{flushright}
\vspace*{15mm}
\begin{center}
{\Large \bf Hyperon polarization from the
twist-3 distribution 
in the unpolarized proton-proton collision}
\vspace{1.5cm}\\
 {\sc Yuji Koike$^1$, Kenta Yabe$^2$ and Shinsuke Yoshida$^1$}
\\[0.7cm]
\vspace*{0.1cm}
{\it $^1$ Department of Physics, Niigata University, Ikarashi, Niigata 950-2181, Japan}

\vspace{0.2cm}

{\it $^2$ Graduate School of Science and Technology, Niigata University,
Ikarashi, Niigata 950-2181, Japan}\\[3cm]

{\large \bf Abstract} \end{center}
\vspace{0.2cm}

\noindent
We investigate the transverse polarization of a hyperon produced in the 
high-energy unpolarized proton-proton
collision, $pp\to\Lambda^\uparrow X$,
based on the collinear twist-3 factorization formalism.  
We focus on the contribution from the twist-3 distribution in one of the unpolarized
proton and the transversity fragmentation function for the final hyperon.
Utilizing the ``master formula" for the soft-gluon-pole cross section,
we clarify the reason for why it receives only the derivative term of the twist-3 distribution.  
We also present the first computation of the soft-fermion-pole cross section and found that it vanishes.  
This means that the derivative of the soft-gluon-pole function is the only source for
the twist-3 cross section from the unpolarized twist-3 distribution, which provides a useful
basis for a phenomenological analysis.


\newpage
\section{Introduction}

Understanding the origin of the transverse polarization of a hyperon (hereafter denoted as 
$\Lambda$) produced from the unpolarized proton-proton (proton-nucleus) collision, $pp\to\Lambda^\uparrow X$~\cite{Bunce:1976yb,Smith87, Ho90}, 
has been a big challenge since its first discovery in 1970's, 
along with the left-right asymmetry observed in the pion production, 
$p^\uparrow p\to\pi X$~\cite{Klem:1976ui,Adams:1991rw,Adams:2003fx,Adler:2005in,:2008mi,Adamczyk:2012qj,Adamczyk:2012xd,Adare:2013ekj}.
This is because the conventional framework of perturbative QCD 
led to a negligible effect for these asymmetries~\cite{Kane:1978nd}.  
Since only one hadron in the reaction
is transversely polarized, they are collectively called transverse single-spin-asymmetries (SSA).  
In the collinear factorization of perturbative QCD,
the SSA appears as a twist-3 observable~\cite{Efremov:1981sh}, and thus it was necessary to
explore the formalism beyond twist-2 for a proper description of SSAs.  
Extensive study 
has been performed along this line and the collinear twist-3 formalism has been well 
established~\cite{Qiu:1991pp,Eguchi:2006qz,Eguchi:2006mc,Koike:2006qv,Yuan:2009dw,Metz:2012ct,Kanazawa:2013uia,Beppu:2010qn}.  
While there has been many works on the SSA in 
$p^\uparrow p\to \pi X$~\cite{Metz:2012ct,Qiu:1998ia,
Kanazawa:2000hz,Kanazawa:2000kp,Kouvaris:2006zy,Koike:2007rq,Koike:2009ge,
Kanazawa:2010au,Kanazawa:2011bg,Kang:2011hk,Kang:2012xf,Kang:2010zzb,Beppu:2013uda,Kanazawa:2014dca}, 
there are only a few works on the 
hyperon polarization in $pp$ collision~\cite{Kanazawa:2001a,Zhou:2008}.  
We will address this issue based on the collinear twist-3 formalism.

Twist-3 effects are, in general, represented in terms of multi-parton (quark-gluon and purely gluonic) correlations
in the distribution and fragmentation functions.  
In particular, the real twist-3 distributions
give rise to the cross section through the pole part of an internal propagator in the hard 
scattering part, which supplies the phase necessary to cause naively $T$-odd SSAs.
For SSAs in the inclusive hadron production in the hadron-hadron collision,
those poles are classified as the soft-gluon-pole (SGP) and the soft-fermion-pole (SFP).  
For the SGP contribution, a convenient ``master formula" was 
invented~\cite{Koike:2006qv,Koike:2007rq,Koike:2011a,Koike:2011b}, 
which reduces the corresponding partonic hard scattering part to a 2-parton$\to$2-parton
scattering cross section, and simplifies the actual calculation greatly.  
Owing to this master formula, the structure of the SGP cross section has become transparent.
For example, 
for the case of $p^\uparrow p\to\{\pi,\gamma\} X$, 
the formula clarified~\cite{Koike:2006qv,Koike:2007rq} why
the {\it total} partonic hard part for the SGP contribution
from the twist-3 quark-gluon correlation function in the polarized nucleon 
becomes the same as the twist-2 unpolarized partonic cross section for $pp\to\pi X$ 
except for the kinematic and color factors,
and why the SGP function $G_F(x,x)$ appears in the combination of
$dG_F(x,x)/dx-G_F(x,x)$ in the cross section, which was found in \cite{Kouvaris:2006zy} by a direct calculation.  
The SFP contribution for the above processes has also been derived~\cite{Koike:2009ge,Kanazawa:2011er}, and
its potential importance was also explored for $p^\uparrow p\to\pi X$~\cite{Kanazawa:2010au,Kanazawa:2011bg}.  
In addition to the quark-gluon correlation in the polarized nucleon, 
this process receives other twist-3 contributions and extensive works have been performed for the
multi-gluon correlation
in the polarized nucleon~\cite{Beppu:2013uda,Koike:2011nx}, the twist-3 distribution in the unpolarized nucleon~\cite{Kanazawa:2000hz,Kanazawa:2000kp}
and the twist-3 fragmentation function for $\pi$~\cite{Metz:2012ct,Kanazawa:2014dca}.

Under this circumstance, we will discuss, in this paper, the 
contribution
from the twist-3 quark-gluon correlation function $E_F(x_1,x_2)$ in the unpolarized nucleon
to the hyperon polarization in $pp\to \Lambda^\uparrow X$.  
This twist-3 distribution is chiral-odd and is combined with
the chiral-odd transversity fragmentation function for $\Lambda^\uparrow$.  
As another source for this observable, twist-3 fragmentation function for the 
final $\Lambda^\uparrow$ also contributes, on which we shall work in a future 
study.\footnote{In \cite{Kanazawa:2015jxa}, the complete twist-3 cross section including the twist-3 fragmentation contribution
for the leptoproduction of the polarized $\Lambda$, $ep\to\Lambda^\uparrow X$, has been derived.}
The purpose of this paper is twofold:
We first present the rederivation of the SGP cross section 
in terms of the master formula.  The SGP cross section, in general, consists of the
derivative and nonderivative terms.  The former was first calculated in \cite{Kanazawa:2001a} 
and the latter was shown to vanish in \cite{Zhou:2008}, which is different from the case of
$p^\uparrow p\to\{\pi,\gamma\} X$.  We will clarify why this happens in the light of the master formula.  
Next we present the first calculation of the SFP contribution to
complete the cross section from the twist-3 distribution. 
As in $p^\uparrow p\to\pi X$ the effect of SFP has a potential importance to the cross section.
However, this contribution turns out to vanish after summing all the diagrams.  

The remainder of the paper is organized as follows: 
in Sec.\,2, we introduce the nonperturbative functions relevant to the present study. 
In Sec.\,3, we present the rederivation of the SGP cross section using
the master formula for the twist-3 distribution contribution to $pp\to \Lambda^{\uparrow}X$.  
In Sec.\,4, we present the first calculation for the SFP contribution.
Sec.\,5 is devoted to a brief summary of the present work.


\section{Twist-3 distribution and the transversity fragmentation function}

We consider the inclusive production of a transversely polarized spin-1/2 hyperon (represented by $\Lambda$) 
from the unpolarized proton-proton collision, 
\beq
p(p)+p(p')\to \Lambda^{\uparrow}(P_h,S_{\perp})+X, 
\eeq
where $p$, $p'$ and $P_h$ represent the momenta of each hadron
and $S_\perp$ is the polarization vector for $\Lambda$.
In the framework of the collinear factorization, 
this hyperon polarization 
can be generated by the two types of mechanisms, i.e., the effect of
the twist-3 distribution function in one of the 
unpolarized protons and that of the 
twist-3 fragmentation function for the polarized $\Lambda$.
In this paper, we focus on the former contribution.

We first introduce the nonperturbative functions relevant to our study. 
The twist-2  unpolarized quark and gluon densities,  $f^q_1(x)$ and 
$f^g_1(x)$, are defined as 
\beq
&&\int{d\lambda\over 2\pi}e^{i\lambda x}
\la p|\bar{\psi}_j(0)\psi_i(\lambda n)|p\ra ={1\over 2}(\slash{p})_{ij}f^q_1(x)+\cdots, 
\label{twist2q}\\
&&\int{d\lambda\over 2\pi}e^{i\lambda x}\la p|
F^{\alpha n}(0)F^{\beta n}(\lambda n)|p\ra
=-{1\over 2}xf^g_1(x)g_{\perp}^{\alpha\beta}(p)+\cdots,
\label{twist2g}
\eeq
where $\psi_i$ is the quark field with spinor index $i$ and 
the projection tensor is defined as 
$g_{\perp}^{\alpha\beta}(p)\equiv g^{\alpha\beta}-p^{\alpha}n^{\beta}-p^{\beta}n^{\alpha}$
with the light-like vector satisfying $p\cdot n=1$.  
$F^{\alpha n}\equiv F^{\alpha\gamma}n_{\gamma}$
represents gluon's 
field strength tensor.
Here and below we suppress the gauge link operators in the distribution and fragmentation functions
for simplicity.  

The twist-3 distribution function $E_F(x_1,x_2)$ in the unpolarized nucleon  
is defined as\,\cite{Kanazawa:2000hz,Kanazawa:2001a}
\beq
&&\int{d\lambda\over 2\pi}\int{d\mu\over 2\pi}
e^{i\lambda x_1}e^{i\mu(x_2-x_1)}\la p|\bar{\psi}_j(0)gF^{\alpha n}(\mu n)
\psi_i(\lambda n)|p\ra \nonumber\\
&&\qquad\qquad= {M_N\over 4}\epsilon^{\alpha\beta np}(\gamma_5\gamma_{\beta}\slash{p})_{ij}
E_F(x_1,x_2)+\cdots,
\label{F-type}
\eeq
where we introduced the nucleon mass $M_N$ to define the function $E_F(x_1,x_2)$ as dimensionless.
From hermiticity and $PT$-invariance, 
$E_F(x_1,x_2)$ is 
real and symmetric as  
$E_F(x_1,x_2)=E_F(x_2,x_1)$.
Another twist-3 distribution function 
${E}_{D}(x_1,x_2)$ obtained 
by replacing  $gF^{\alpha w}(\mu n)$ in (\ref{F-type}) 
by the covariant derivative 
$D^{\alpha}({\mu n})=\partial^{\alpha}-igA^{\alpha}(\mu n)$
is related to $E_F(x_1,x_2)$ as
\beq
E_{D}(x_1,x_2)={\cal P}\left({1\over x_1-x_2}\right)E_{F}(x_1,x_2),
\label{GIR}
\eeq
where ${\cal P}$ indicates the principal value.  
This relation shows that $E_F(x_1,x_2)$ is the only independent twist-3 quark-gluon correlation function
in the unpolarized nucleon.   

The twist-2 fragmentation function $H_1(z)$ for the polarized $\Lambda$  is 
defined as~\cite{Kanazawa:2001a}
\beq
{1\over N}\sum_{X}\int{d\lambda\over 2\pi}e^{i{\lambda\over z}}
\la 0|\psi_i(\lambda w)|\Lambda(P_h,S_{\perp})\ X\ra\la \Lambda(P_h,S_{\perp})\ X|\bar{\psi}_j(0)|0\ra
=(\gamma_5\slash{S}_{\perp}{\slash{p}_c})_{ij}H_1(z)\cdots,
\label{twist2-frag}
\eeq
where $w^{\mu}$ is a light-like vector satisfying $P_h\cdot w=1$, and $p_c\equiv P_h/z$ is the momentum 
of the quark fragmenting into $\Lambda^\uparrow$. 
The combination of the two chiral-odd functions $E_F(x_1,x_2)$ and $H_1(z)$ can generate the
transverse polarization of the hyperon.


\section{SGP contribution to $p+p\to \Lambda^{\uparrow}+X$ }

According to the general formalism of the twist-3 calculation for SSAs~\cite{Eguchi:2006mc},
the SGP component of a twist-3 distribution function in the nucleon contributes as derivative and nonderivative
terms.  
For the case of the SGP contribution of $p^\uparrow p\to\{\pi,\gamma\} X$, it was shown that
the derivative and the nonderivative terms have the common partonic hard cross sections
and thus the total SGP function contributes to the cross section in
the particular combination.  The origin of this simplification was clearly understood
in terms of the ``master formula" which shows the {\it total} SGP partonic hard cross section
is connected to the twist-2 unpolarized cross section for $pp\to\{\pi,\gamma\} X$ with different color
factors~\cite{Koike:2006qv,Koike:2007rq}.  
For the case of $pp\to \Lambda^{\uparrow}X$, the derivative term was first calculated in \cite{Kanazawa:2001a} and the  
nonderivative terms was also calculated in \cite{Zhou:2008} and was shown to vanish, indicating that
the SGP contribution is the sole contribution.
Here we calculate the SGP cross section for $pp\to \Lambda^{\uparrow}X$
in the light of the master formula and clarify the origin for the vanishing
nonderivative terms.  We emphasize that the master formula itself is valid for this process as well, 
while its outcome differs from the case of $p^\uparrow p\to\{\pi,\gamma\} X$, and this difference
does not imply the ``violation" of the master formula as claimed in \cite{Zhou:2008}.

Applying the formalism developed in \cite{Eguchi:2006mc}, one can obtain the
SGP contribution to $pp\to \Lambda^{\uparrow}X$
from the following formula:
\beq
E_{P_h}{d\Delta\sigma\over d^3P_h}&=&{iM_N\over 64\pi^2 s}
\int{dx'\over x'}f_1(x')\int{dz\over z^2}H_1(z) \nonumber\\
&&\times\left.
\int {dx_1\over x_1}\int dx_2 E_F(x_1,x_2)\epsilon^{\alpha\beta np}
{\partial\over \partial k_2^{\alpha}}\left(S^{\rm I}_{\lambda\beta}(k_1,k_2)p^\lambda
+S^{\rm F}_{\lambda\beta}(k_1,k_2)p^\lambda\right)
\right|_{k_i=x_ip},
\label{twist3cross}
\eeq
where $s=(p+p')^2$ is the square of the center of mass energy, and the partonic 
hard parts 
$S_{\lambda\beta}^{\rm I,F}(k_1,k_2)p^\lambda$ 
are obtained from diagrams shown in Fig. 1 by taking the spinor- and color- traces 
with the appropriate projections for the distribution and fragmentation functions
in (\ref{twist2q}), (\ref{twist2g}), (\ref{F-type}) and (\ref{twist2-frag})\footnote{
Since we included an extra $x_1$ into $S_{\lambda\beta}^{I,F}$ to project out $E_F(x_1,x_2)$, the integration measure
became $dx_1/x_1$.}.  
$S_{\lambda\beta}^{\rm I}$ and $S_{\lambda\beta}^{\rm F}$ correspond to
the initial-state-interaction (ISI) and the final-state-interaction (FSI), respectively.  
In $S_{\lambda\beta}^{\rm I,F}(k_1,k_2)$, the Lorentz index $\lambda$ corresponds to
that for the coherent gluon line with the momentum $k_2-k_1$ in Fig. 1,
and $\beta$ is for the projection of $E_F(x_1,x_2)$ in (\ref{F-type}).  
The SGP contribution was obtained by a direct computation without referring to the master formula 
in \cite{Kanazawa:2001a,Zhou:2008}. 
\begin{figure}[b]
\begin{center}
  \includegraphics[height=3cm,width=12cm]{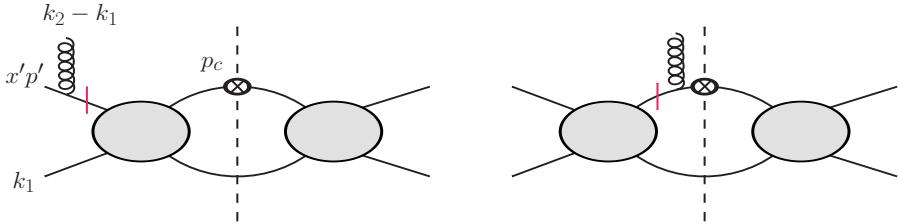}\hspace{1cm}
\end{center}
 \caption{
Diagrammatic representation of the hard part for the SGP contribution.  
Left (Right) diagram corresponds to ISI (FSI) and gives 
$S^{\rm I}_{\lambda\beta}(k_1,k_2)$ ($S^{\rm F}_{\lambda\beta}(k_1,k_2)$). 
The SGP is given as a pole part of the bared propagator.  The circled cross indicates the fragmentation insertion for $\Lambda^{\uparrow}$.
 Each blob represents the $2\to 2$ scattering amplitude.}
\end{figure} 
\begin{figure}[th]
\begin{center}
  \includegraphics[height=3cm,width=7cm]{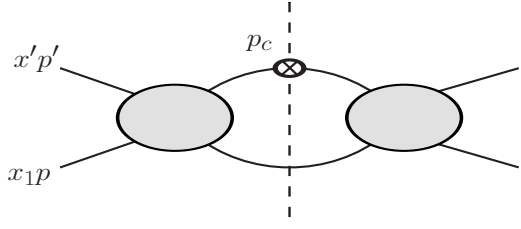}\hspace{1cm}
\end{center}
 \caption{Diagrammatic representation for $\widetilde{S}^{\rm I,F}_{\beta\gamma}(x_1p,x'p',p_c).$}
\end{figure} 
Here we take an alternative approach based on the master formula
which reduces the SGP cross section to a 2$\to$2 partonic cross section.  
Following the procedure described in \cite{Koike:2006qv,Koike:2007rq,Koike:2011a,Koike:2011b,Kanazawa:2014nea}, one obtains
the hard part as
\beq
\hspace{-1cm}
&&\left. \epsilon^{\alpha\beta np} {\partial   S^{\rm I}_{\lambda\beta}(k_1,k_2)p^\lambda \over \partial k_2^{\alpha}}  
\right|_{k_i=x_ip}^{\rm SGP}=
\left[{-1\over x_2-x_1+i\epsilon}
\right]^{\rm pole}
\epsilon^{\alpha\beta np}
S^{\gamma}_{\perp}{d\over d(x'p'^{\alpha})}
\widetilde{S}^{\rm I}_{\beta\gamma}(x_1p,x'p',p_c), 
\label{initial} \\
\hspace{-1cm}&&\left. \epsilon^{\alpha\beta np}
{\partial  S^{\rm F}_{\lambda\beta}(k_1,k_2)p^\lambda \over \partial k_2^{\alpha}}
\right|_{k_i=x_ip}^{\rm SGP}
=\left[{1\over x_1-x_2+i\epsilon}\right]^{\rm pole}
\epsilon^{\alpha\beta np}\nonumber\\
&&\qquad\qquad\qquad\qquad\qquad \times\left[
S^{\gamma}_{\perp}{d\over dp^{\alpha}_c}
+{1\over p\cdot p_c}\left\{(p\cdot S_{\perp})g_{\alpha}^{\ \gamma}
-S_{\perp\alpha}p^{\gamma}\right\}\right]\widetilde{S}^{{\rm F}}_{\beta\gamma}(x_1p,x'p',p_c), 
\label{final}
\eeq
where
$\epsilon^{\alpha\beta np}
S^{\gamma}_\perp
\widetilde{S}^{\rm I}_{\beta\gamma}(x_1p,x'p',p_c)$ 
represents the partonic cross section for the 2$\to$2 scattering process, $q(x_1p)+b(x'p')\to q(p_c) + q(x_1p+x'p'-p_c)$ ($b=q$ or $g$), 
as shown in Fig. 2.  It is obtained 
by the spinor projection $\epsilon^{\alpha\beta np}\gamma_5\gamma_\beta x_1\pslash$ 
for the initial parton with momentum $x_1p$, the unpolarized projection for the parton $b$ with momentum $x'p'$, and the projection
$\gamma_5\Sslash_\perp\pslash_c$ for the final parton fragmenting into the polarized $\Lambda$, 
but has the same color factor as the ISI diagrams 
in Fig. 1.\footnote{We have factored out the spin vector $S_\perp^\gamma$ to define $\widetilde{S}^{\rm I}_{\beta\gamma}(x_1p,x'p',p_c)$ 
for later convenience.}
Except for the color factor, 
the twist-2 partonic cross section for
the spin-transfer reaction 
$p^\uparrow(p,S_{N\perp}) + p(p')\to \Lambda^\uparrow(P_h,S_\perp)+X$~\cite{Stratmann:1992gu} can be written as
$S_{N\perp}^{\beta}S_\perp^\gamma \widetilde{S}^{\rm I}_{\beta\gamma}(x_1p,x'p',p_c)$.  
$\widetilde{S}^{\rm F}_{\beta\gamma}(x_1p,x'p',p_c)$ in (\ref{final})
is defined similarly but with the color factors for the FSI diagrams in Fig. 1, and thus
$\widetilde{S}^{\rm I}_{\beta\gamma}$ and 
$\widetilde{S}^{\rm F}_{\beta\gamma}$ defer only in the color factors.
In taking the derivative ${d\over dx'p'^\alpha}$ in (\ref{initial}), the on-shell form $p'^\mu=\left(p'^+={\vec{p'_\perp}^2 \over 2p'^-}, p'^-, \vec{p'_\perp}\right)$
needs to be used.  
Likewise the derivative ${d\over dp_c^\alpha}$ needs to be taken with 
$p_c^\mu=\left({\vec{p_{c_\perp}}^2 \over 2p_c^-}, p_c^-, \vec{p_{c_\perp}}\right)$ in (\ref{final}).  
We note the appearance of the extra terms in
(\ref{final}) for FSI compared with
(\ref{initial}) for ISI, which is different from the case of 
the $G_F(x,x)$ contribution to $p^\uparrow p\to\pi X$.

To calculate the derivatives ${d\widetilde{S}^{\rm I}_{\beta\gamma}\over dx'p'^\alpha}$ and ${d\widetilde{S}^{\rm F}_{\beta\gamma}\over dp_c^\alpha}$
in (\ref{initial}) and (\ref{final}), 
we note that 
$\widetilde{S}^{\rm Y}_{\beta\gamma}(xp,x'p',p_c)$ (Y=I,\ F) can be expanded as
\beq
\widetilde{S}^{\rm Y}_{\beta\gamma}(xp,x'p',p_c)&=&a^{\rm Y}_1g_{\beta\gamma}+a^{\rm Y}_2 xp_{\beta} xp_{\gamma}
+a^{\rm Y}_3 x'p'_{\beta} x'p'_{\gamma} 
+a^{\rm Y}_4   p_{c\beta} p_{c\gamma}
+a^{\rm Y}_5   x'p'_{\beta}  xp_{\gamma} \nonumber\\
&+&a^{\rm Y}_6xp_{\beta}x'p'_{\gamma}
+a^{\rm Y}_7 p_{c\beta} xp_{\gamma} 
+a^{\rm Y}_8 xp_{\beta} p_{c\gamma}
+a^{\rm Y}_9 p_{c\beta} x'p'_{\gamma} 
+a^{\rm Y}_{10} x'p'_{\beta}p_{c\gamma}, 
\label{expansion}
\eeq
where 
each coefficient $a^{\rm Y}_i=a^{\rm Y}_i(\hat{s},\hat{t},\hat{u})$ ($i=1,\cdots,10$) is a scalar function of the Mandelstam variables, 
$\hat{s}=(xp+x'p')^2$,
$\hat{t}=(xp-p_c)^2$ and
$\hat{u}=(x'p'-p_c)^2$.  
We further introduce the partonic cross sections $\hat{\sigma}^{\rm Y}_i(\hat{s},\hat{t},\hat{u})$
(Y=I, F) by the relation
\beq
a^{\rm Y}_i(\hat{s},\hat{t},\hat{u})=\hat{\sigma}^{\rm Y}_i(\hat{s},\hat{t},\hat{u})\delta(\hat{s}+\hat{t}+\hat{u}),
\eeq
for later use.  Here we work in a frame in which $p$ and $p'$ are collinear.  
The derivative $d/dx'p'^\alpha$ and $d/dp_c^\alpha$ on $a^{\rm Y}_i$ can be performed through
$\hat{u}$ to obtain\footnote{In taking the derivative with respect to
$p'^\alpha_\perp$, we first keep $p'_\perp\neq 0$ to take the derivative and then put $p'_\perp=0$.}
\beq
{d\over d(x'p'^{\alpha})}a^{\rm I}_i(\hat{s},\hat{t},\hat{u})
=-2p_{c{\alpha}}{\partial\over \partial\hat{u}}a^{\rm I}_i(\hat{s},\hat{t},\hat{u}), \\
{d\over dp_c^{\alpha}}a^{\rm F}_i(\hat{s},\hat{t},\hat{u})
=-2\Bigl({\hat{s}\over \hat{t}}\Bigr)p_{c{\alpha}}{\partial\over \partial\hat{u}}a^{\rm F}_i(\hat{s},\hat{t},\hat{u}).
\eeq
Using (\ref{initial}) and  (\ref{expansion}) in (\ref{twist3cross}), one can express the cross section
from the ISI in terms of $a^{\rm I}_i$ ($i=1,\cdots,10$).  It is easy to see that
only $a_{1,9}^{\rm I}$ survive and one obtains
\beq
&&E_{P_h}{d\Delta\sigma^{\rm ISI}\over d^3P_h}
=-{M_N\over 64\pi s}
\int{dx'\over x'}f_1(x')\int{dz\over z^2}H_1(z)\int {dx\over x} E_F(x,x)
\epsilon^{\alpha\beta np}S_{\perp}^{\gamma}{d\over d(x'p'^{\alpha})}
\widetilde{S}^{\rm I}_{\beta\gamma}(xp,x'p',p_c)\nonumber\\
& &\quad =-{M_N\over 32\pi s}
\int{dx'\over x'}f_1(x')\int{dz\over z^2}H_1(z)\int {dx\over x}E_F(x,x)
\epsilon^{p_cpnS_{\perp}}({\partial\over \partial\hat{u}}a^{\rm I}_1+{1\over 2}a^{\rm I}_9)\nonumber\\
& &\quad =-{M_N\over 32\pi s}
\int{dx'\over x'}f_1(x')\int{dz\over z^2}H_1(z)
\int {dx\over x}\,\delta\left(\hat{s}+\hat{t}+\hat{u}\right)\,\epsilon^{p_cpnS_{\perp}} \nonumber\\
&&\qquad\qquad\qquad\qquad\qquad \times\Bigl[x{d E_F(x,x) \over dx}{1\over \hat{u}}\hat{\sigma}^{\rm I}_1
+E_F(x,x)(-{1\over \hat{u}}\hat{\sigma}^{\rm I}_1+{1\over 2}\hat{\sigma}^{\rm I}_9)\Bigr],
\label{ISIcross}
\eeq
where we used the relation
$\left( \hat{s}{\partial\over \partial\hat{s}}
+\hat{t}{\partial\over \partial\hat{t}}
+\hat{u}{\partial\over \partial\hat{u}}\right)
\hat{\sigma}^{\rm Y}_1(\hat{s},\hat{t},\hat{u})=0$
which follows from the scale invariance property
$\hat{\sigma}^{\rm Y}_1(\lambda\hat{s},\lambda\hat{t},\lambda\hat{u})=\hat{\sigma}^{\rm Y}_1(\hat{s},\hat{t},\hat{u})$.  

Repeating the same steps for the FSI contribution, one obtains the corresponding cross section as
\beq
E_{P_h}{d\Delta\sigma^{\rm FSI}\over d^3P_h}&=&{M_N\over 32\pi s}
\int{dx'\over x'}f_1(x')\int{dz\over z^2}H_1(z)
\int {dx\over x}\delta(\hat{s}+\hat{t}+\hat{u})\epsilon^{p_cpnS_{\perp}} \nonumber\\
&&\times\left[x{d E_F(x,x) \over dx} 
\Bigl({\hat{s}\over \hat{t}\hat{u}}\Bigr)\hat{\sigma}^{\rm F}_1
+E_F(x,x)\Bigl({\hat{s}\over \hat{t}}\Bigr)
(-{1\over \hat{u}}\hat{\sigma}^{\rm F}_1+{1\over 2}\hat{\sigma}^{\rm F}_9)\right],
\label{FSIcross}
\eeq
which is formally the same form as the ISI contribution (\ref{ISIcross})
except for the kinematic factor $\hat{s}/\hat{t}$.  
From (\ref{ISIcross}) and (\ref{FSIcross})
one sees that the SGP function does not appear in the 
combination of $x{d\over dx}E_F(x,x)-E_F(x,x)$ if
$\hat{\sigma}^{\rm I,F}_9\neq 0$.  
In fact we found, by the direct calculation, the relation 
$\hat{\sigma}^{\rm I,F}_9=2\hat{\sigma}^{\rm I,F}_1/\hat{u}$ 
in all the channels.  
This means that only the derivative of the SGP function 
contributes to
$pp\to \Lambda^{\uparrow} X$, which was found by \cite{Zhou:2008}.  
This is in contrast to the case of $p^{\uparrow}p\to \{\pi,\gamma\} X$~\cite{Koike:2007rq},
where the SGP function appears in the combination of 
$x{d\over dx}G_F(x,x)-G_F(x,x)$.  
This way 
the final form of the SGP contribution to $pp\to \Lambda^{\uparrow} X$
is obtained as follows~\cite{Kanazawa:2001a,Zhou:2008}:
\vspace{3mm}
\beq
&&E_{P_h}{d\Delta\sigma^{\rm SGP}\over d^3P_h} \nonumber\\
&&\quad={\pi M_N\alpha_s^2\over s}\epsilon^{P_hpnS_{\perp}}\sum_{a,b,c}
\int{dx'\over x'}f^b_1(x')\int{dz\over z^3}H^c_1(z)\int d x \, {d   E^a_F(x,x)  \over dx}\,
\sigma_{ab\to c}\,\delta(\hat{s}+\hat{t}+\hat{u}),\hspace{2mm}
\eeq
where the partonic cross section in each channel is given by
\beq
\sigma_{qq'\to q}&=&{1\over N^2}{2\hat{s}\over \hat{t}^2}-{1\over N^2}{\hat{s}^2\over \hat{t}^3},
\hspace{5mm}\sigma_{qq\to q}=
\sigma_{qq'\to q}-\Bigl({1\over N}+{1\over N^3}\Bigr){\hat{s}\over \hat{t}\hat{u}}
+{1\over N^3}{\hat{s}^2\over \hat{t}^2\hat{u}}, \nonumber\\
\sigma_{q\bar{q}'\to q}&=&\Bigl({N^2-2\over N^2}\Bigr){\hat{s}\over \hat{t}^2}
-{1\over N^2}{\hat{s}^2\over \hat{t}^3},
\hspace{5mm}\sigma_{q\bar{q}\to q}=\sigma_{q\bar{q}'\to q\bar{q}'}
+{1\over N^3}{1\over t}+{1\over N^3}{\hat{s}\over \hat{t}^2}, \nonumber\\
\sigma_{q\bar{q}\to \bar{q}}&=&-{1\over N^3}{1\over \hat{u}}
+\Bigl({1\over N}+{1\over N^3}\Bigr){\hat{s}\over \hat{t}\hat{u}}, \nonumber\\
\sigma_{qg\to q}&=&-{N^2\over N^2-1}{\hat{u}\over t^2}+{1\over N^2-1}{1\over \hat{u}}
-{1\over N^2(N^2-1)}{\hat{s}\over \hat{t}\hat{u}}-{1\over (N^2-1)}
{2\hat{s}^2\over \hat{t}^3},
\eeq
where $N=3$ is the number of colors for a quark.  
Summarizing this section, 
we have shown that the partonic hard cross section for the SGP contribution 
from the twist-3 distribution to  
$pp\to \Lambda^{\uparrow} X$
can also be reduced to a certain 2$\to$2 scattering cross section
(``master formula"),
and the relation can be conveniently used to derive the cross section.  
We have also shown that
the particular relation for the above 2$\to$2 scattering cross section
leads to the result that only the derivative term with $dE_F(x,x)/dx$ appears in the SGP 
cross section for $pp\to \Lambda^{\uparrow} X$, which is in contrast the case of
$p^\uparrow p\to\{\pi,\gamma\} X$ where the SGP function appears through the combination
$x{d\over dx}G_F(x,x)-G_F(x,x)$.


\section{SFP contribution to $p+p\to \Lambda^{\uparrow}+X$ }

In this section we present the first derivation for the SFP contribution from the twist-3 unpolarized quark distribution.  
According to the formalism of \cite{Eguchi:2006mc}, it receives only the nonderivative contribution and is given by
\beq
E_{P_h}{d\Delta\sigma\over d^3P_h}&=&{iM_N\over 64\pi^2s}
\int{dx'\over x'}f_1(x')\int{dz\over z^2}H_1(z) \nonumber\\
&&\times\int dx_1\int dx_2E_F(x_1,x_2)
\epsilon^{\alpha\beta np}{\cal P}\Bigl({1\over x_1-x_2}\Bigr)S^{\rm SFP}_{\alpha\beta}(x_1p,x_2p), 
\eeq
where $S^{\rm SFP}_{\alpha\beta}(x_1p,x_2p)$ represents the hard part which is obtained by calculating the diagrams 
shown in Figs. 3-5.  The meaning of the Lorentz indices 
is the same as $S_{\lambda\beta}^{\rm I,F}(k_1,k_2)$ in (\ref{twist3cross}).  
By the direct calculation of the diagrams, it turned out that the SFP contribution
completely vanishes in all channels after summing over all the 
diagrams.\footnote{The SFP contribution from $E_F$ to $p^\uparrow p\to\gamma X$ was
also shown to vanish in \cite{Kanazawa:2014nea}.}
As a result, only the derivative term
of SGP contribution survives in the final cross section formula.
\begin{figure}[h]
\begin{center}
  \includegraphics[height=3cm,width=12cm]{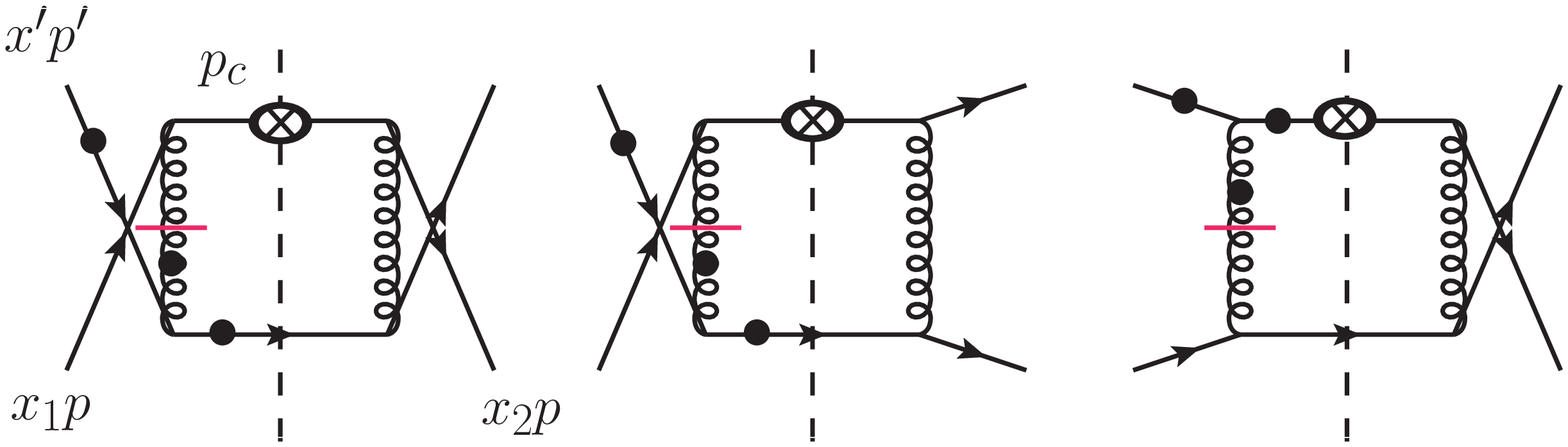}\hspace{1cm}
\end{center}
\begin{center}
  \includegraphics[height=3cm,width=14cm]{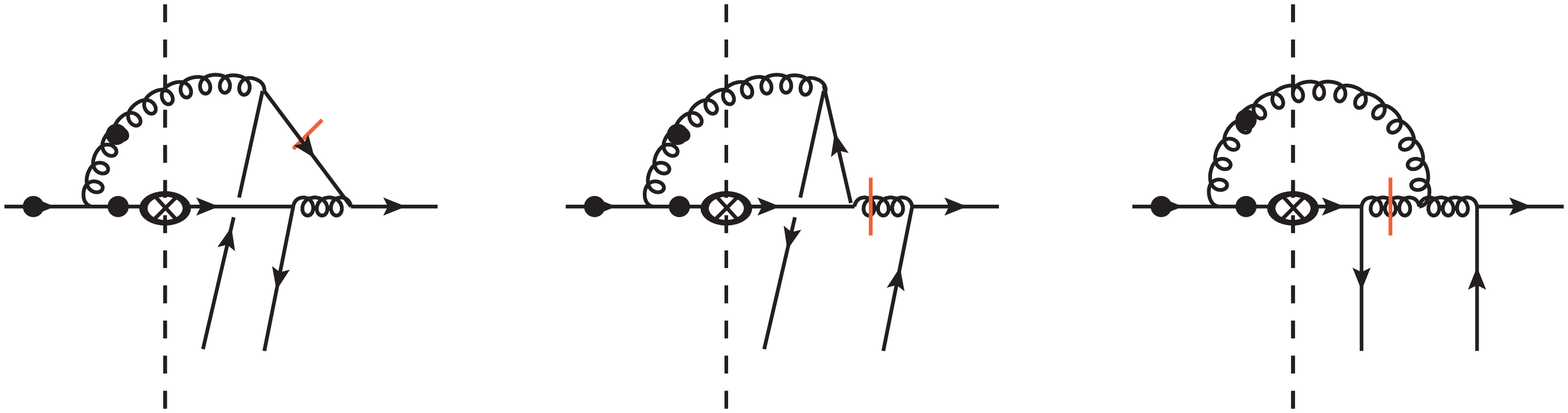}\hspace{1cm}
\end{center}
 \caption{Lowest order diagrams for the hard part of the SFP contribution in the $qq'\to qq'$ and $qq\to qq$ channels.  
 The twist-3 distribution contributes from the lower side of each diagram.  
 For each diagram, three diagrams corresponding to a different attachment of the coherent gluon line 
 to one of the dots need to be considered.  
The barred propagator gives rise to SFP.  Mirror diagrams should also be included. }
\end{figure} 
\begin{figure}[h]
\begin{center}
  \includegraphics[height=3cm,width=12cm]{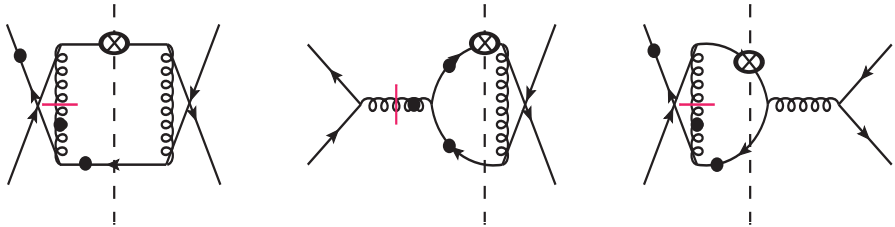}\hspace{1cm}
\end{center}
 \caption{The same as Fig. 3, but for the $q\bar{q}\to q\bar{q}$ channel. 
Diagrams for the $\bar{q} q\to q\bar{q}$ channel are obtained by reversing the arrows of the quark lines
and shifting the fragmentation insertion to the other quark line crossing the final-state cut.}
\end{figure}
\begin{figure}[th]
\begin{center}
  \includegraphics[height=9cm,width=12cm]{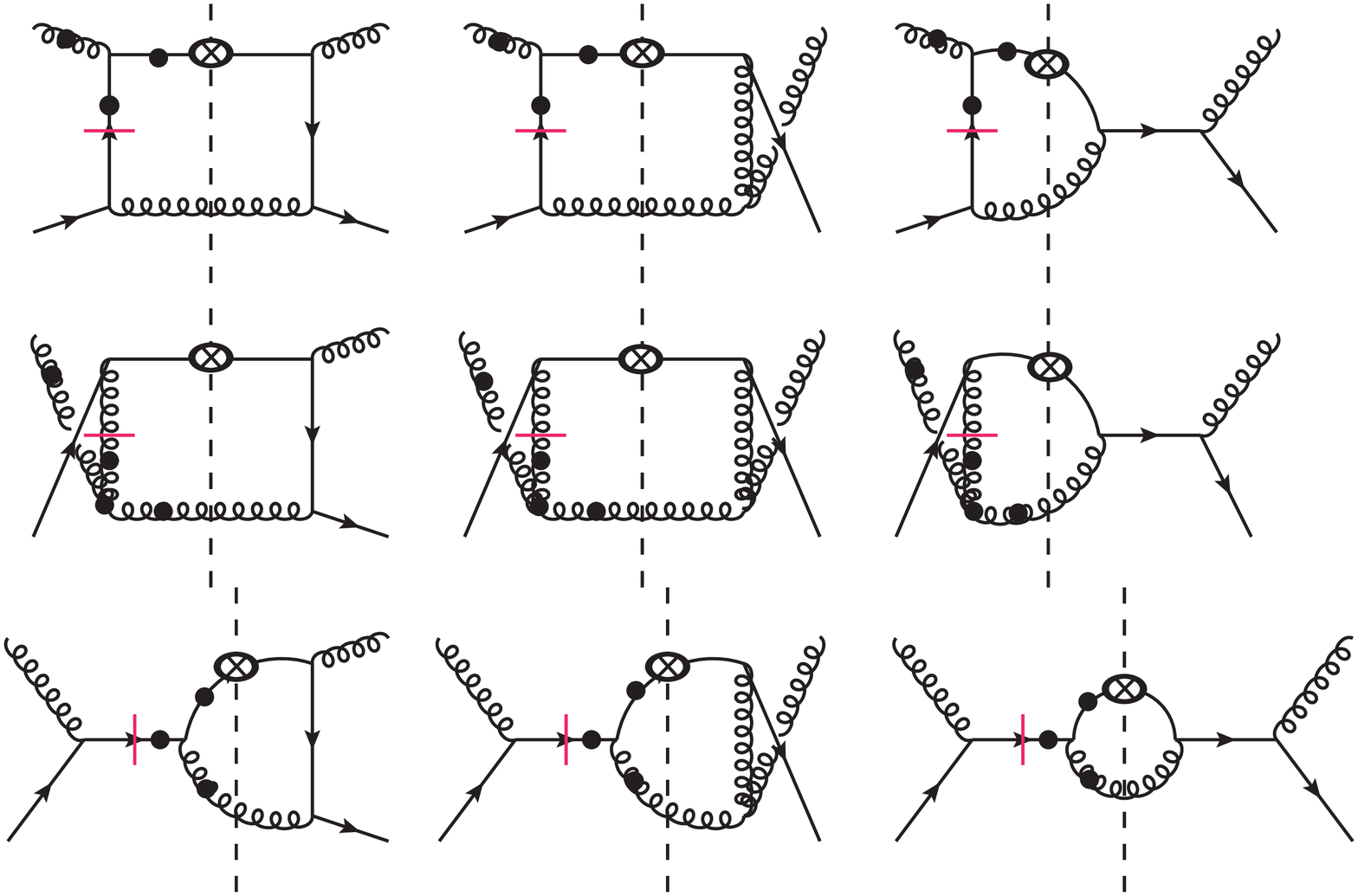}\hspace{1cm}
\end{center}
\begin{center}
  \includegraphics[height=3cm,width=14cm]{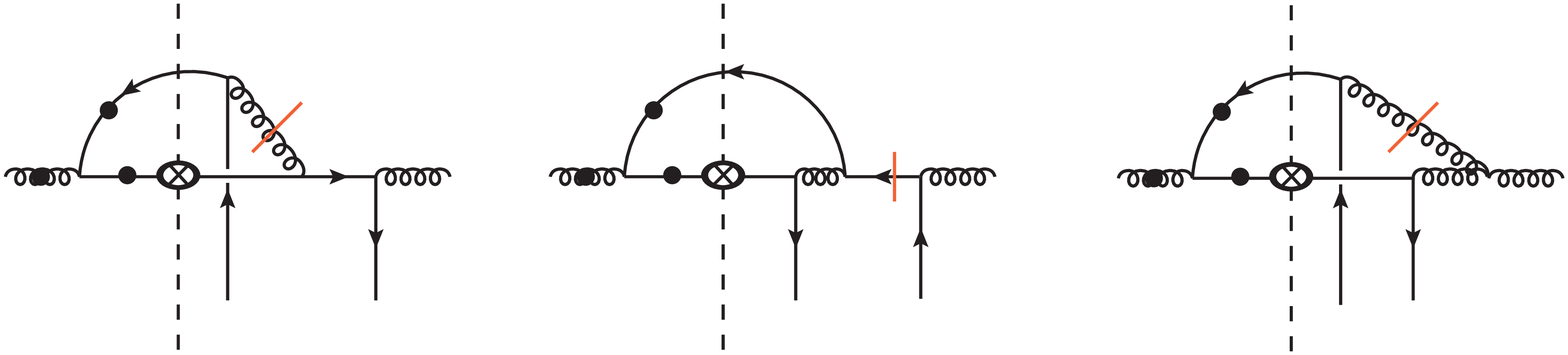}\hspace{1cm}
\end{center}
 \caption{The same as Fig.3, but for the $qg\to qg$ channel.}
\end{figure}

\section{Summary}

In this paper we have studied the transverse polarization of a hyperon produced in the 
unpolarized proton-proton
collision, $pp\to\Lambda^\uparrow X$.  
This is an example of the single spin asymmetry (SSA) in high-energy inclusive reactions
and has been a long-standing issue since its discovery in 1970's.
In perturbative QCD, this phenomenon occurs as a twist-3 effect in the framework of the collinear factorization.  
We have focused on the effect of the 
twist-3 quark-gluon correlation function in one of the initial proton
combined with the transversity fragmentation function
for the final hyperon.
In principle, the cross section appears as the SGP and SFP contributions,
and we presented the full derivation for those contributions.  
The SGP cross section was derived some time ago~\cite{Kanazawa:2001a,Zhou:2008},
which showed only the derivative of the SGP function $dE_F(x,x)/dx$ contributes,
unlike the case for the SSA in $p^\uparrow p\to \{\pi,\gamma\} X$.  
We presented a rederivation of the SGP cross section in the light of the ``master formula"~\cite{Koike:2006qv,Koike:2007rq}
and have clarified the origin of the difference between the two processes.  
We also calculated the SFP cross section for the first time and showed that
it vanishes in all channels after summing over all diagrams.  
We conclude from this study  
that the twist-3 effect in the unpolarized proton is embodied
solely through the derivative of the SGP function, $d E_F(x,x)/dx$, in $pp\to \Lambda^{\uparrow}X$, 
which provides a useful basis for future phenomenological study on the hyperon polarization.  
For the complete understanding of the origin of the hyperon polarization, however, 
we need to include the twist-3 fragmentation function for the final $\Lambda^{\uparrow}$.
We will address this issue in a future publication. 

\section*{Acknowledgments}

This work has been supported by the Grant-in-Aid for
Scientific Research from the Japanese Society of Promotion of Science
under Contract No.~26287040 (Y.K. and S.Y.).

\end{document}